\documentclass[twocolumn]{aastex7}

\usepackage{CJK}
\usepackage{amsmath}

\newcommand{\cta}{\citetalias}
\defcitealias{Lu_TRACE}{L24}

\begin{document}
\begin{CJK*}{UTF8}{gbsn}
\title{A Coordinate System for Dynamical Instabilities in Hierarchical Systems in REBOUND}

\author[0000-0003-0834-8645]{Tiger Lu (陆均)}
\altaffiliation{Flatiron Research Fellow}
\affiliation{Center for Computational Astrophysics, Flatiron Institute, 162 5th Avenue, New York, NY 10010, USA}
\email[show]{tlu@flatironinstitute.org}  

\author[0000-0002-9354-3551]{Garett Brown}
\affiliation{Independent Researcher}
\email{astro@gbrown.ca}

\begin{abstract}
We implement coordinates suitable for studying wide binary systems in \texttt{TRACE}, a hybrid integrator in the widely used open-source \textit{N}-body integration package \texttt{REBOUND}. This is a regime in which traditional hybrid integrators perform poorly. The coordinate system supports close encounters between any pair of bodies in the system. We describe the implementation of this coordinate system and benchmark its performance against other integrators in the \texttt{REBOUND} ecosystem. In tests of planet-planet scattering, stellar flybys, and ZLK oscillations \texttt{TRACE} in wide binary coordinates is qualitatively correct when other hybrid methods fail, and in many cases returns statistically similar results to the high-precision \texttt{IAS15} integrator with up to $9\times$ speedups. We also provide some guidelines for when use of these coordinates are appropriate.
\end{abstract}

\section{Introduction}
The \textit{N}-body problem is one of the oldest unsolved problems in classical mechanics -- despite centuries of effort, no general reasonable analytic solution exists for $N \geq 3$ \citep{poincare1890equations, sundman1913memoire} and numerical simulations are often the only viable approach for studying the evolution of planetary systems. This comes at a steep computational cost: in a planetary system the shortest orbital periods are often measured in days, while the dynamical processes that sculpt their architectures may unfold over gigayears. Bridging these twelve or more orders of magnitude in timescale is the central numerical challenge of the field.

For the so-called planetary \textit{N}-body problem, characterized by a dominant central ``star" orbited by many smaller ``planets", \cite{wisdom_holman_1991} and \cite{kinoshita1991symplectic} developed an efficient, accurate and widely used algorithm known today as the ``Wisdom-Holman" method, which enabled integration of planetary systems on gigayear timescales. The genius of the Wisdom-Holman method was in the realization that it was possible to split the Hamiltonian governing the equations of planetary motion into a dominant part, corresponding to Keplerian motion about the host star, and a much smaller part that incorporates effects from other planets in the systems as minor perturbations to this dominant Keplerian motion. This allows Wisdom-Holman integrators to take large timesteps without sacrificing accuracy, and for this excellent blend of long-term accuracy and speed Wisdom-Holman integrators have become the gold standard when studying the long-term evolution of planetary systems.

The specifics of how the Hamiltonian is split depend on the choice of coordinate system used. While Wisdom-Holman methods can in principle be formulated in any canonical coordinate system (see \citealt{rein_tamayo_brown_2019} for a review), arguably the two most widely adopted are Jacobi coordinates and Democratic Heliocentric coordinates \citep[DHC;][]{duncan_levison_lee_1998}, each with well-documented strengths and weaknesses that have been the subject of considerable discussion in the literature \citep{hernandez_2017, ReinTamayo2019, rein2026dhc, hernandez2026dhc}. In brief: Jacobi coordinates exactly solve the two-body problem and hence are more accurate. However, Jacobi coordinates are inflexible in that they are inherently tied to a specific orbital ordering. If this ordering ever changes -- for instance, due to dynamical instabilities -- Jacobi coordinates cannot be used. In such cases, one must stomach the slightly less accurate DHC in order to support the possibility of orbit crossings. As such, DHC is the coordinate system of choice for so-called hybrid integrators such as \texttt{MERCURY} \citep{chambers1998making}, \texttt{SyMBA} \citep{duncan_levison_lee_1998, levison_duncan_2000}, \texttt{GENGA} \citep{grimm2014genga}, \texttt{MERCURIUS} \citep{rein_2019}, and \texttt{TRACE} \citep[hereafter L24]{Lu_TRACE} which seek to leverage the accuracy benefits of Wisdom-Holman methods over long timescales while occasionally switching over to more flexible integration techniques when close encounters call for it.

\begin{figure}
    \centering
    \includegraphics[width=0.42\textwidth]{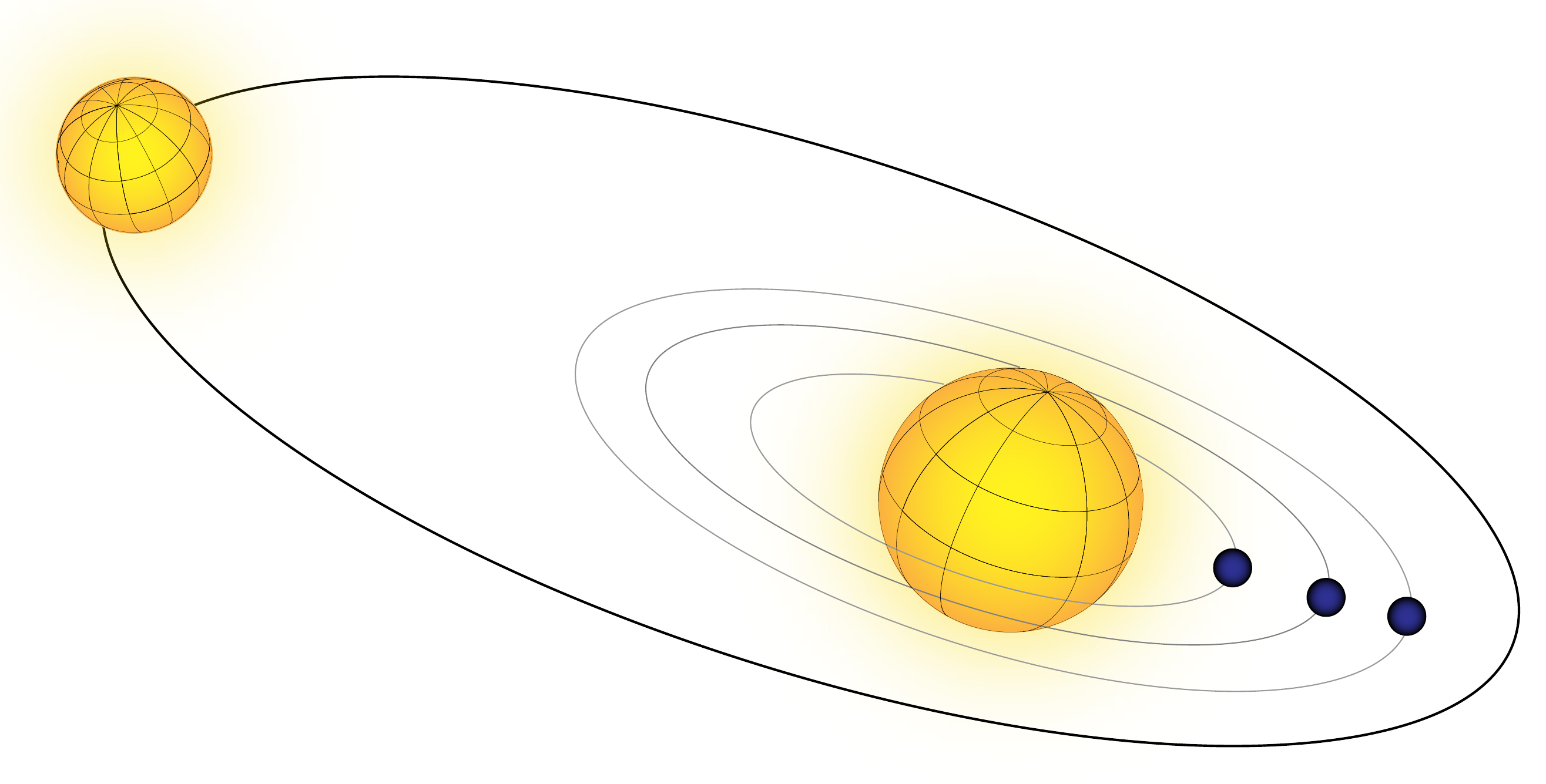}
    \caption{Schematic representation of a wide binary system, where a system of planets orbits around one component of a binary star system. Orbits and physical sizes not to scale.}
    \label{fig:schematic}
\end{figure}

Wide binary systems, in which a planetary system orbits one component of a stellar binary (see Figure \ref{fig:schematic}), would seem to violate the fundamental assumption Wisdom-Holman methods are built upon: that of a dominant central mass. Wisdom-Holman methods in Jacobi coordinates handle such systems surprisingly well in practice, but indeed perform poorly in DHC. We thus aim to address a specific system archetype in this work: binary star systems in which DHC fails, but also involve close encounters and orbital instabilities that render Jacobi coordinates unusable. Understanding dynamical instabilities in binary systems has immense scientific value: intriguing clues pointing to the profound influence binary companions have on planetary system architectures and demographics have surfaced in recent years \citep[e.g.][]{kaib2013disruption, ngo2016friends, christian2025wide, sullivan2026architectures}.

Fortuitously this problem was solved by \cite{chambers2002wb}, who presented a modification of DHC suitable for binary systems. These coordinates were included as an option in the \texttt{MERCURY} integrator. In this work, we implement these coordinates as an option in the \texttt{TRACE} hybrid integrator, which is available in the open source C/Python package \texttt{REBOUND}\footnote{\url{https://rebound.readthedocs.io/}} \citep{Rein_2012}. Close encounters between pairs of planets, as well as between planets and either star, are supported.

The plan for this paper is as follows. In \S \ref{sec:wbmap} we review the coordinate system derived by \cite{chambers2002wb}, and detail the resulting modifications to the \texttt{TRACE} algorithm. In \S \ref{sec:paramspace} we provide guidelines as to what regions of parameter space these coordinates are a significant improvement over DHC. In \S \ref{sec:tests} we apply the integrator to several astrophysically relevant systems and benchmark its performance against other integrators in the \texttt{REBOUND} ecosystem. We conclude in \S \ref{sec:conclusion}.

\section{The Wide Binary Map}
\label{sec:wbmap}
The Hamiltonian for a planetary system of \textit{N} planets (subscripted $i=1,\dots,N$) orbiting one star of a binary system (subscripted A, hereafter simply ``star") can be written:

\begin{equation}
\label{eq:ham}
\begin{split}
    \mathcal{H} = & \frac{p_\mathrm{A}^2}{2 m_\mathrm{A}} + \frac{p_\mathrm{B}^2}{2 m_\mathrm{B}} + \sum_{i=1}^N \frac{p_i^2}{2 m_i} \\
    & - G \bigg[\frac{m_\mathrm{A} m_\mathrm{B}}{|\vec{q}_\mathrm{A} - \vec{q}_\mathrm{B}|} + m_\mathrm{A} \sum_{i=1}^N \frac{m_i}{|\vec{q}_\mathrm{A} - \vec{q}_\mathrm{i}|}\\
    & + m_\mathrm{B} \sum_{i=1}^N \frac{m_i}{|\vec{q}_\mathrm{B} - \vec{q}_\mathrm{i}|} + \sum_{i=1}^N \sum_{j>i} \frac{m_i m_j}{|\vec{q}_i - \vec{q}_j|} \bigg],
\end{split}
\end{equation}
where $\vec{q}, \vec{p}$ are the canonical coordinates/momenta and $m$ the masses. The companion star (hereafter simply ``companion") is denoted with the subscript B.

\subsection{Wide Binary Coordinates}
\label{sec:coordinates}
Equation \ref{eq:ham} is very difficult to solve numerically. \cite{chambers2002wb} leveraged the hierarchical nature of binary systems to define a new coordinate system. These coordinates, originally coined ``Wide Binary Coordinates" and hereafter simply \texttt{WB}, are defined:

\begin{equation}
\begin{split} 
    \vec{Q}_\mathrm{A} & = \frac{m_\mathrm{A}\vec{q}_\mathrm{A} + m_\mathrm{B}\vec{q}_\mathrm{B} + \sum_jm_j\vec{q}_j}{m_\mathrm{tot}}, \\
    \vec{Q}_i & = \vec{q}_i - \vec{q}_\mathrm{A},\\
    \vec{Q}_\mathrm{B} & = \vec{q}_B - \frac{m_\mathrm{A}\vec{q}_\mathrm{A} +  \sum_jm_j\vec{q}_j}{m_\mathrm{tot}}, \\
\end{split}
\end{equation}
and momenta defined:

\begin{equation}
    \begin{split}
        \vec{P}_\mathrm{A} & = \vec{p}_\mathrm{A} + \vec{p}_\mathrm{B} + \sum_{j}\vec{p}_j, \\
        \vec{P}_i & = \vec{p}_i - m_i\frac{\vec{p}_\mathrm{A} + \sum_{j}\vec{p}_j}{m_A + \sum_{j}m_j}, \\
        \vec{P}_\mathrm{B} & = \vec{p}_\mathrm{B} + m_\mathrm{B}\frac{\vec{p}_\mathrm{A} + \vec{p}_\mathrm{B} + \sum_{j}\vec{p}_j}{m_\mathrm{tot}},\\
    \end{split}
\end{equation}
where $m_\mathrm{tot} = m_\mathrm{A} + m_\mathrm{B} + \sum_jm_j$, and each summation over $j$ is assumed to run from $j=1$ to $j=N$. In these coordinates, Equation \ref{eq:ham} may be rewritten

\begin{equation}
\label{eq:wbham}
    \mathcal{H} = \mathcal{H}_\mathrm{Jump} + \mathcal{H}_\mathrm{Interaction} + \mathcal{H}_\mathrm{Kepler}
\end{equation}
where $\mathcal{H}_\mathrm{Jump}$ describes the barycentric motion of the central star:

\begin{equation}
\label{eq:hjump}
    \mathcal{H}_\mathrm{Jump} = \frac{1}{2m_\mathrm{A}} \left| \sum_{i=1}^{N} \vec{P}_i \right|^2,
\end{equation}
$\mathcal{H}_\mathrm{Interaction}$ describes both planet-planet and planet-companion interactions:

\begin{equation}
\label{eq:hint}
    \begin{split}
        \mathcal{H}_\mathrm{Interaction} = & \underbrace{Gm_\mathrm{B} m_\mathrm{A} \left( \frac{1}{Q_\mathrm{B}} - \frac{1}{|\vec{Q}_\mathrm{B} + \vec{s}|} \right)}_\mathrm{star-companion \: interaction} \\ & + \underbrace{Gm_\mathrm{B} \sum_{i=1}^{N} m_i \left( \frac{1}{Q_\mathrm{B}} - \frac{1}{|\vec{Q}_\mathrm{B} - \vec{Q}_i + \vec{s}|} \right)}_\mathrm{planet-companion \: interactions} \\ 
        & - \underbrace{\sum_{i=1}^{N} \sum_{j>i} \frac{Gm_i m_j}{|\vec{Q}_j - \vec{Q}_i|}}_{\mathrm{planet-planet \: interactions}},
    \end{split}
\end{equation}
where $\vec{s}$ is the center of mass of the planetary system, relative to the star:

\begin{equation}
    \vec{s} \equiv \frac{\sum_{i=1}^Nm_i\vec{Q}_i}{m_\mathrm{A}+\sum_{i=1}^Nm_i}.
\end{equation}
Finally, $\mathcal{H}_\mathrm{Kepler}$ describes the Keplerian motion (of both the planets and the companion) about the star:

\begin{equation}
\label{eq:hkep}
\begin{split}
    \mathcal{H}_\mathrm{Kepler} = & \underbrace{\frac{P_\mathrm{B}^2}{2\mu_{\text{bin}}} - \frac{Gm_{\text{tot}}\mu_{\text{bin}}}{Q_\mathrm{B}}}_\mathrm{companion \: orbit} \\
    & + \underbrace{\sum_{i=1}^{N} \left( \frac{P_i^2}{2m_i} - \frac{Gm_\mathrm{A} m_i}{Q_i} \right)}_\mathrm{planet \: orbits},
\end{split}
\end{equation}
where $\mu_\mathrm{bin}\equiv(m_\mathrm{A}+\sum_jm_j) m_\mathrm{B}/m_\mathrm{tot}$ is the reduced mass of the system. Note that in the absence of a companion, these coordinates reduce to the familiar \texttt{DHC}. The equations of motion governed by the sub-Hamiltonians $\mathcal{H}_\mathrm{Interaction}$ and $\mathcal{H}_\mathrm{Jump}$ both admit trivial analytic solutions, while $\mathcal{H}_\mathrm{Kepler}$ corresponds to Kepler's equation and can be solved with a Kepler solver. For a timestep $h$, one timestep of the \texttt{WB} integrator consists of the following substeps:

\begin{enumerate}
    \item Advance $\mathcal{H}_\mathrm{Interaction}$ for $h/2$,  \label{splitting}
    \item Advance $\mathcal{H}_\mathrm{Jump}$ for $h/2$,
    \item Advance $\mathcal{H}_\mathrm{Kepler}$ for $h$,
    \item Advance $\mathcal{H}_\mathrm{Jump}$ for $h/2$,
    \item Advance $\mathcal{H}_\mathrm{Interaction}$ for $h/2$.
\end{enumerate}
The idea of this splitting, as in all Wisdom-Holman schemes, is for the Keplerian motion $\mathcal{H}_\mathrm{Kepler}$ to dominate the system's dynamics, with $\mathcal{H}_\mathrm{Interaction}$ and $\mathcal{H}_\mathrm{Jump}$ acting only as small perturbations.

\subsection{Switching Conditions}
The assumption of a dominant $\mathcal{H}_\mathrm{Kepler}$ breaks down in the regime of close encounters, either between pairs of planets or between a planet and one of the stars. The approach adopted by many hybrid integrators is to move terms from the other sub-Hamiltonians into $\mathcal{H}_\mathrm{Kepler}$ as necessary to ensure it always remains dominant over the other sub-Hamiltonians. When this happens $\mathcal{H}_\mathrm{Kepler}$ no longer exactly corresponds to Kepler's equation and thus a Kepler solver cannot be used, but a traditional integrator such as \texttt{IAS15} \citep{rein_ias15} or Bulirsch-Stoer \citep{press02} may be used to solve the resulting equations of motion.

The main novelty of this work is the application of the time-reversible switching scheme of \texttt{TRACE} to \texttt{WB} coordinates. All encounters -- between pairs of planets, between planets and the star, and between planets and the companion -- are supported. We provide a general qualitative description of these switching conditions here, and refer interested readers to \cta{Lu_TRACE} for a more in-depth review. One timestep of the \texttt{WB} integrator proceeds as such:

\begin{enumerate}
    \item We first check for encounters between planets and either star. The terms associated with these interactions are coupled, so there is no convenient way to shift certain terms to $\mathcal{H}_\mathrm{Kepler}$ here. Hence, our approach is to convert our system back to inertial coordinates and solve Equation \ref{eq:ham} if one of these encounters is detected. By default \texttt{REBOUND}'s implementation of Bulirsch-Stoer \citep{Lu_2023} is used, but the user may also opt for \texttt{IAS15}.

    \begin{itemize}
        \item Close approaches with the star are flagged with the same approach used in \cta{Lu_TRACE}, using a criterion derived by \cite{pham2024new} involving ratios of multiple higher-order derivatives.
        \item Close approaches with the companion are flagged if any planet comes within some fraction $R_\mathrm{crit,WB}$ of the companion's Hill radius. From our testing we found $R_\mathrm{crit,WB} = 0.1$ to be reasonable and this set as the default. This value can be adjusted by the user via the \texttt{ri\_trace.r\_crit\_WB} field.
    \end{itemize}

    \item If no encounter between planets and stars are detected, we next check for close encounters between pairs of planets. We integrate Equation \ref{eq:wbham} with the procedure described in \S \ref{sec:coordinates}. However, the sub-Hamiltonians $\mathcal{H}_\mathrm{Interaction}$ and $\mathcal{H}_\mathrm{Kepler}$ are now modulated by the switching terms

    \begin{equation}
    \begin{split}
        \mathcal{H}_\mathrm{Interaction} = & \:Gm_\mathrm{B} m_\mathrm{A} \left( \frac{1}{Q_\mathrm{B}} - \frac{1}{|\vec{Q}_\mathrm{B} + \vec{s}|} \right)\\ & + Gm_\mathrm{B} \sum_{i=1}^{N} m_i \left( \frac{1}{Q_\mathrm{B}} - \frac{1}{|\vec{Q}_\mathrm{B} - \vec{Q}_i + \vec{s}|} \right) \\ 
        & - \sum_{i=1}^{N} \sum_{j>i} \frac{Gm_i m_j}{|\vec{Q}_j - \vec{Q}_i|} \textcolor{red}{[1-\mathcal{K}_{ij}]},
    \end{split}
    \end{equation}
    
    \begin{equation}
    \begin{split}
        \mathcal{H}_\mathrm{Kepler} = & \frac{P_\mathrm{B}^2}{2\mu_{\text{bin}}} - \frac{Gm_{\text{tot}}\mu_{\text{bin}}}{Q_\mathrm{B}} \\
        & + \sum_{i=1}^{N} \left( \frac{P_i^2}{2m_i} - \frac{Gm_\mathrm{A} m_i}{Q_i} \right)\\
        & - \sum_{i=1}^{N} \sum_{j>i} \frac{Gm_i m_j}{|\vec{Q}_j - \vec{Q}_i|} \textcolor{red}{\mathcal{K}_{ij}},
    \end{split}
    \end{equation}
    where $\mathcal{K}_{ij} = 1$ if there is a close encounter between planets $i$ and $j$, and $\mathcal{K}_{ij} = 0$ otherwise. The criterion for flagging a close encounter between pairs of planets is the same as that of \cta{Lu_TRACE}, which checks for overlapping Hill radii multiplied by some constant \texttt{r\_crit\_hill}. The idea is to move appropriate terms from $\mathcal{H}_\mathrm{Interaction}$ into $\mathcal{H}_\mathrm{Kepler}$ during close encounters such that $\mathcal{H}_\mathrm{Kepler}$ remains the dominant aspect of the motion. The equations of motion associated with $\mathcal{H}_\mathrm{Interaction}$ can always be trivially solved. In the absence of close encounters, those associated with $\mathcal{H}_\mathrm{Kepler}$ are solved with the Kepler solver used in \texttt{WHFast} \citep{rein_2015}, \texttt{REBOUND}'s implementation of the Wisdom-Holman algorithm. If there is a close encounter, the terms in $\mathcal{H}_\mathrm{Kepler}$ associated with that close encounter are solved with Bulirsch-Stoer.

    \item At the end of a timestep, all close encounters are re-checked and the timestep is potentially rejected and repeated with the time-reversible algorithm of \cite{hernandez_2023} implemented in \texttt{TRACE}.
    
\end{enumerate}

\section{Parameter Space Study}
\label{sec:paramspace}
Here we explore the parameter space in which \texttt{WB} coordinates offer a meaningful advantage over \texttt{DHC}. The results of this section represent only a test of the coordinates themselves, with further tests of close encounters in \texttt{WB} coordinates discussed in \S \ref{sec:tests}.

We conducted a suite of simulations, all of which included the Sun, Jupiter and Saturn on their present-day orbits. In each simulation, we also added a companion on a circular orbit. We sampled the mass of the companion from a log-uniform distribution from $1M_\mathrm{J}$ to $3M_\odot$, and its semimajor axis uniformly from $50$ to $1000$ AU. Each parameter was sampled at $100$ values, yielding 
$10,000$ simulations per suite in total. We integrated each system for $10^6$ Jupiter orbits, using a timestep of $1/20$th Jupiter's initial orbital period. We ran one suite with \texttt{TRACE} in \texttt{WB} coordinates, and one in \texttt{DHC}\footnote{The default switching condition in \texttt{TRACE} flags a close encounter if particles pass within $3$ Hill radii of one another. This is physically reasonable for planets, but not a binary companion for which this condition is far too generous -- if this default condition were used, every timestep would be flagged as a close encounter and integrated with Bulirsch-Stoer. While this would indeed correctly integrate the system, it would be far slower and would not represent a fair comparison of the coordinate systems. Hence, for this and all subsequent tests, we manually set the switching criterion for the binary to it's Hill radius multiplied by \texttt{r\_crit\_WB} when \texttt{TRACE DHC} is used.}. All systems remained stable with no close encounters throughout each integration. The mean energy errors were $3.53 \times 10^{-6}$ and $3.23 \times 10^{-4}$ for \texttt{WB} and \texttt{DHC}, respectively.

\begin{figure}
    \centering
    \includegraphics[width=0.48\textwidth]{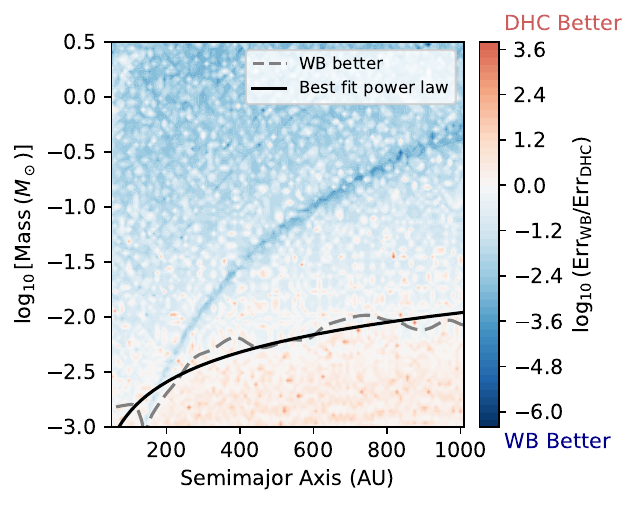}
    \caption{Comparison of energy errors in a parameter space of binary companions, in semimajor axis-mass space. For each companion, the ratio of energy errors between an integration with \texttt{WB} coordinates and an integration with \texttt{DHC} coordinates is plotted, both with the \texttt{TRACE} integrator. Blue corresponds to \texttt{WB} performing better, and red the opposite. The gray contour roughly corresponds to the limit where \texttt{WB} outperforms \texttt{DHC}, and this is reasonably well fit by Equation \ref{eq:plawfit}.}
    \label{fig:paramspace}
\end{figure}
Our results are shown in Figure \ref{fig:paramspace}. For each binary companion, we plot the ratio of the energy error $|E_\mathrm{final} - E_\mathrm{initial}/E_\mathrm{initial}|$ for the \texttt{WB} integration over the \texttt{DHC} integration. Blue corresponds to regions in parameter space where \texttt{WB} outperforms \texttt{DHC}, and red the opposite. \texttt{WB} coordinates perform significantly better for larger, more distant companions. A gray dashed contour roughly corresponding to where \texttt{WB} and \texttt{DHC} perform equivalently is shown. This contour is reasonably well fit by a power law

\begin{equation}
\label{eq:plawfit}
    \log_{10} \left(\frac{M}{M_\odot}\right) = 0.9 \log_{10} \left(\frac{a}{\text{1 AU}}\right) - 4.67,
\end{equation}
which we offer as a guideline for when \texttt{WB} coordinates offer a significant advantage over \texttt{DHC}. \texttt{WB} coordinates are strongly preferred for the majority of true stellar companions.

We also ran the simulations above with \texttt{WHFAST} in Jacobi coordinates. These yielded an average energy error of $1.92 \times 10^{-6}$. \texttt{WB} coordinates were competitive with \texttt{WHFAST} across the entirety of the explored parameter space. We also note a dark blue band in Figure \ref{fig:paramspace} where \texttt{TRACE WB} performs up to six orders of magnitude better than \texttt{TRACE DHC}. It is unclear what the physical interpretation of this band is, and further investigation is warranted -- but we note that this feature is due to \texttt{TRACE WB} performing extremely well, not \texttt{TRACE DHC} performing poorly. The same spike in performance is seen in \texttt{WHFAST}.

\section{Tests}
\label{sec:tests}
In this section, we apply \texttt{TRACE} in \texttt{WB} coordinates to some astrophysically relevant test problems, and benchmark its performance against other integrators in \texttt{REBOUND}. All integrators use their default, out-of-the-box settings, other than the switching criterion modification to \texttt{TRACE DHC} described in \S \ref{sec:paramspace}.

\subsection{Planet-Planet Scattering}
Planet-planet scattering is believed to have been a ubiquitous process in the late stages of planet formation \citep[e.g.][]{davies_2004, juric_tremaine_2008, chatterjee_2008}, and was envisioned as one of the most powerful use cases of \texttt{TRACE}. We repeat a similar experiment to the one performed in Section 5.3 of \cta{Lu_TRACE}. The original experiment consists of three Jupiter-like planets around a sun-like star. The Jupiters are initialized in compact orbits such that they rapidly experience dynamical instabilities. In this experiment, we also add a $0.5 M_\odot$ star on a circular, $200$ AU orbit. As such, this experiment tests both planet-planet encounters, planet-star encounters, and the robustness of the \texttt{WB} coordinates in general. We have run $500$ simulations each with \texttt{IAS15} and \texttt{TRACE} with both \texttt{WB} and \texttt{DHC} coordinates.

\begin{figure*}
    \centering
    \includegraphics{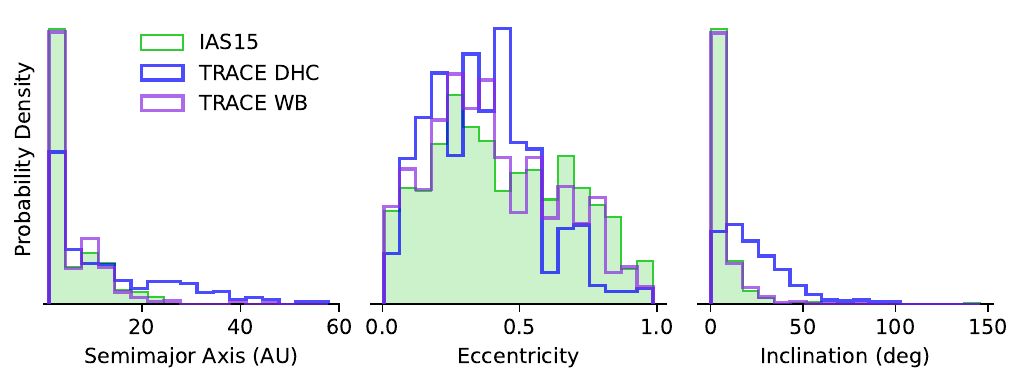}
    \caption{Distributions of semimajor axis, eccentricity, and inclination from results of three-body scattering experiments including a binary perturber. Results from \texttt{IAS15} are plotted as a solid green histogram, and taken to be the gold standard for numerical integration. Results from \texttt{TRACE DHC} and \texttt{TRACE WB} are plotted as blue and purple step histograms, respectively.}
    \label{fig:elements}
\end{figure*}

We report results from our simulations in Figure \ref{fig:elements}. We plot histograms of semimajor axis, eccentricity, and inclination of all surviving planets. There are noticeable differences between \texttt{IAS15} and \texttt{TRACE} \texttt{DHC}, which has a tendency to overpredict semimajor axis and inclination while underpredicting eccentricity. \texttt{TRACE} \texttt{WB} is significantly more consistent with the \texttt{IAS15} demographics.

For the \texttt{IAS15}, \texttt{TRACE DHC} and \texttt{TRACE WB} suites, the mean energy errors are $10^{-9.39}, 10^{-1.14}$ and $10^{-2.62}$, respectively. The average runtimes are $55.2$, $3.34$ and $5.82$ minutes, respectively. In a regime where \texttt{TRACE DHC} performs poorly, \texttt{TRACE WB} recovers statistically similar results to \texttt{IAS15} simulations while boasting over a $9\times$ speedup.

\subsection{Stellar Flybys}
Planetary systems do not exist in a vacuum, and stellar flybys can dramatically shape the architectures of planetary systems \citep[e.g.][]{Spurzem2009, malmberg2011flyby, parker2012dynamical, li2015cross, Cai+2017, SchoettlerOwen2024}. \texttt{TRACE} in \texttt{DHC} coordinates does not resolve these encounters well.

We used the \texttt{AIRBALL} package \citep{airball} to simulate the effect of stellar flybys on a planetary system\footnote{The setup of our simulations closely follows one of the examples in the \texttt{AIRBALL} documentation: \url{https://airball.gbrown.ca/examples/adiabatic-tests/}}. Our planetary system is initialized with a sun-like star and a Neptune-like planet. We performed 300 flyby simulations each with \texttt{IAS15} and \texttt{TRACE} in both \texttt{DHC} and \texttt{WB}. The flyby star is initialized at a distance of $10^5$ AU from the host star and we fix its pericenter approach to $16.5\times$ Neptune's orbit. The orientation and velocity of the flyby is randomized. The mass of the stellar companion varies from a Jupiter mass to $100$ Solar masses. A timestep is selected using the \texttt{timestep\_for\_perihelion\_resolution} function in \texttt{AIRBALL}, which selects a timestep sufficient to resolve a highly eccentric orbit based on the criterion of \cite{wisdom_2015} and \cite{hernandez_2022}. The flyby duration is of order 0.5 Myr.

\begin{figure}
    \centering
    \includegraphics{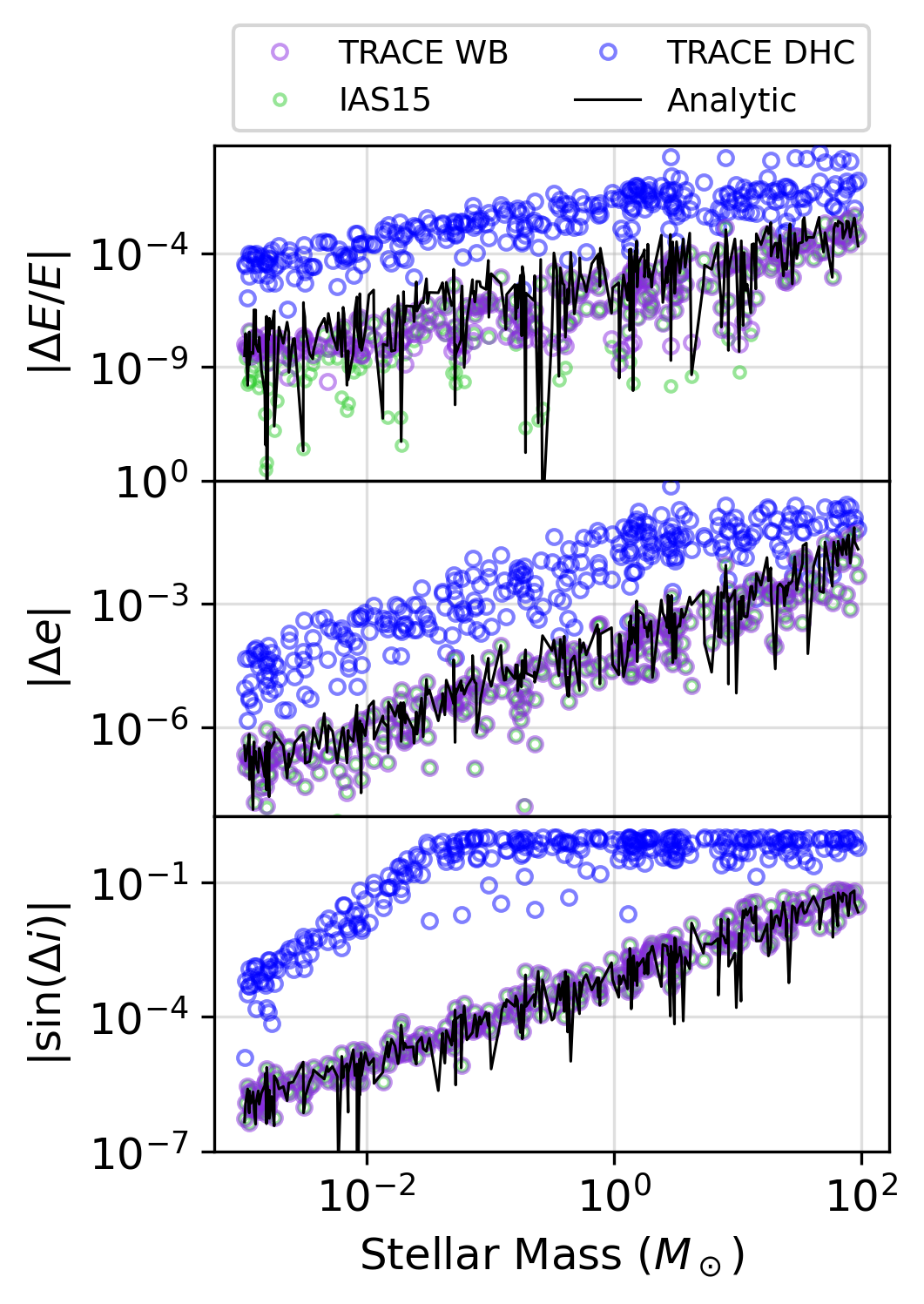}
    \caption{Effect of a stellar flyby as a function of perturber mass on a single planet system's energy error \textbf{(top)}, eccentricity excitation \textbf{(middle)} and inclination excitation \textbf{(bottom)}. We compare results from \texttt{IAS15} (green), \texttt{TRACE WB} (violet) and \texttt{TRACE DHC} (blue) to analytic predictions (black).}
    \label{fig:airball}
\end{figure}

Figure \ref{fig:airball} shows the results of our simulations. We plot relative change in the specific orbital energy of Neptune as well as change in Neptune's eccentricity and inclination as a function of stellar mass, and compare with analytic predictions. The analytic predictions were derived by \cite{Heggie1975, Heggie1996, malmberg2011flyby} for the energy, eccentricity, and inclination estimates, respectively \cite[see also][]{Rickman1976, RoyHaddow2003, Heggie2006, Spurzem2009}. The estimates focus primarily on the interactions of three similarly sized point masses (usually ``stars''), but \cite{Spurzem2009} discuss the applications to hierarchical systems that have a dominant central mass and a smaller orbiting body. From our simulations, \texttt{TRACE WB} and \texttt{IAS15} both reproduce analytic predictions well, while \texttt{TRACE DHC} fails across the board. The mean runtimes for \texttt{IAS15}, \texttt{TRACE WB} and \texttt{TRACE DHC} were 0.34, 0.15 and 0.11 seconds, respectively. \texttt{TRACE WB} boasts a $2\times$ speedup over \texttt{IAS15}, and we stress that the actual integration speedup is likely higher as our simulations' runtimes were dominated by initialization and overhead costs.

We caution that due to the structure of the \texttt{TRACE WB} code, only one flyby at a time is supported. This should not be a major issue -- the effects of stellar flybys on planetary systems follow a L\'{e}vy flight where the overall change to the system is often dominated by a few rare flyby events \citep{BrownRein2022}.

\subsection{ZLK Cycles}
von Zeipel-Lidov-Kozai (ZLK) cycles \citep{von_zeipel_1910, lidov1962evolution, kozai1962secular, naoz2016eccentric} are a well studied phenomenon in hierarchical three-body systems. A planet in the presence of a massive inclined companion will experience coupled eccentricity and inclination oscillations. \cta{Lu_TRACE} identified ZLK cycles as a fail case of \texttt{TRACE} in \texttt{DHC} coordinates.

Generally if numerically integrating ZLK cycles an adaptive timestep integrator such as \texttt{IAS15} is preferred over fixed-timestep integrators. This is because a planet undergoing ZLK cycles experiences epochs of both high and low eccentricity. Accurately resolving the dynamics near pericenter during high-eccentricity phases requires a very small timestep. In particular, \citet{wisdom_2015} show that the timestep  $\tau_p$ should satisfy

\begin{equation}
    \tau_p = \frac{\pi}{8} \sqrt{\frac{(1-e)^3}{1+e} \frac{a^3}{G M_*}}.
\end{equation}
Fixed-timestep integrators face two main difficulties in this context. First, they require prior knowledge of the maximum eccentricity in order to set an appropriate timestep, which is a nontrivial estimate in some regimes \citep[e.g.][]{Li14}. Second, the timestep must remain small throughout the entire integration, even during low-eccentricity phases where such resolution is unnecessary. The speed benefits of fixed-timestep integrators are worth it for systems that only reach moderate eccentricity, but the timesteps required for systems that attain extremely high eccentricity quickly render runtimes untenably long. We note that although \texttt{TRACE} is designed in principle to mitigate this issue via a pericenter switching condition, in practice we found that it can lead to qualitatively different behavior.

We integrated systems consisting of a Sun-like star, a Jupiter-mass planet on an initially circular $5$ AU orbit, and a 0.5$M_\odot$ binary companion on a circular $100$ AU orbit. The binary companion's inclination\footnote{We also ran a set of simulations with $89^\circ$ inclinations, but these proved too computationally costly to complete for the fixed-timestep integrators.} was set to $50^\circ, 60^\circ, 70^\circ$ and $80^\circ$.This setup was run with \texttt{IAS15} and \texttt{TRACE} in both \texttt{WB} and \texttt{DHC} coordinates. For \texttt{TRACE}, we also explored timesteps of $1, 2, 5$ and $10\times \tau_p$ where $\tau_p$ was calculated from the theoretical maximum eccentricity attained for ZLK cycles in the quadrupole limit:

\begin{equation}
    e_\mathrm{max} = \sqrt{1-5/3\cos^2i_\mathrm{mut}}
\end{equation}
where $i_\mathrm{mut}$ is the initial mutual inclination between the companion and the Jupiter. All systems were integrated for $3$ Myr, which is ${\sim}100\times$ the nominal timescale for ZLK cycle in the quadrupole limit \citep{antognini2015timescales}:

\begin{equation}
    \label{eq:zlktimescale}
     t_{\rm ZLK} = \frac{16}{30\pi}\frac{m_\mathrm{A} + m_\mathrm{Jup} + m_\mathrm{B}}{m_\mathrm{B}}\frac{P_\mathrm{comp}^2}{P_\mathrm{Jup}}.
\end{equation}
In practice, the actual number of ZLK oscillations in each simulation was around $10$ and varied with binary inclination. 

\begin{figure*}
    \centering
    \includegraphics{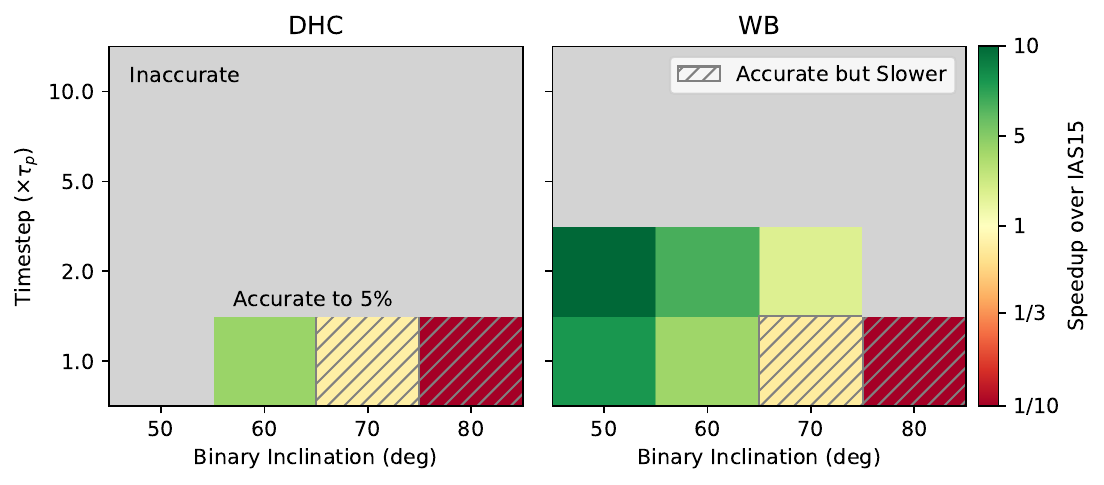}
    \caption{Binary inclination - timestep space in which \texttt{TRACE DHC} and \texttt{WB} are preferred over \texttt{IAS15} for ZLK integrations. Gray squares indicate where \texttt{TRACE} does not reproduce \texttt{IAS15} results. Hatched colored squares indicate where \texttt{TRACE} reproduces \texttt{IAS15} results, but is slower. Unhatched colored squares represent parameter space where \texttt{TRACE} is preferred over \texttt{IAS15}. \texttt{WB} coordinates greatly expand the region of parameter space where \texttt{TRACE} is preferred.}
    \label{fig:zlkcomparison}
\end{figure*}
In Figure \ref{fig:zlkcomparison}, we compare the performance of \texttt{TRACE} in both \texttt{DHC} and \texttt{WB} to \texttt{IAS15}. Each grid cell represents an inclination (directly mapping to maximum eccentricity) -- timestep pairing. For each setup, we first assessed the accuracy of the \texttt{TRACE} simulations. We numerically computed the maximum eccentricity and the mean ZLK oscillation period for each of our simulation suites. If either quantity deviated from the \texttt{IAS15} simulation to $5\%$ (roughly the standard deviation of the \texttt{IAS15} ZLK periods), the cell is shaded gray. Simulations that reproduce \texttt{IAS15} are colored by their runtime relative to \texttt{IAS15}. Gray hatches are overlaid on the simulations where \texttt{IAS15} is faster. Un-hatched colored squares are regions where \texttt{TRACE} would be preferred over \texttt{IAS15}. In \texttt{DHC} coordinates, the preferred parameter space is extremely limited. \texttt{WB} coordinates significantly expand this region. With a sufficiently small timestep, \texttt{TRACE WB} is preferred over \texttt{IAS15} for binary inclinations up to $70^\circ$, corresponding to $e_\mathrm{max} \sim 0.9$. For more extreme eccentricities, the timestep required to resolve pericenter passage becomes so small that \texttt{IAS15}'s adaptive timestep becomes advantageous.

\begin{figure}
    \centering
    \includegraphics{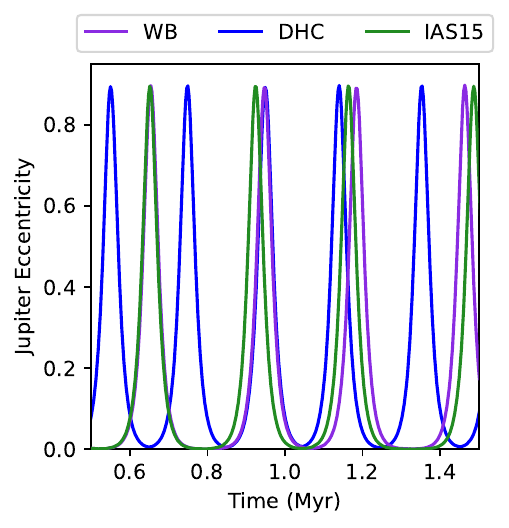}
    \caption{Evolution of Jupiter's eccentricity for a system with a Sun-like star, and Earth-like planet on a 0.1 AU orbit, a Jupiter-like planet on a 5 AU orbit, and a 0.5 $M_\odot$ companion on a 100 AU orbit inclined by $70^\circ$. We compare results from \texttt{IAS15} (green), \texttt{TRACE WB} (purple) and \texttt{TRACE DHC} (blue). For clarity, only a representative slice of the evolution from $0.5$ to $1.5$ Myr is plotted.  \texttt{IAS15} takes ${\sim}20\times$ longer to run than either \texttt{TRACE} run.}
    \label{fig:tslimited}
\end{figure}

One specific but potentially relevant complication is if timestep is restricted for some other reason -- for example, a close-in planet with a companion undergoing ZLK oscillations. In this case, \texttt{IAS15} loses the advantage of an adaptive timestep. We demonstrate this in Figure \ref{fig:tslimited}, where we have run the $i_\mathrm{mut} = 70^\circ$ case, but now include an Earth-like planet on a $0.1$ AU circular orbit. For \texttt{TRACE} we adopt a timestep equal to 1/15th the orbital period of the inner planet, while \texttt{IAS15}'s adaptive timestep remains small in order to resolve the Earth's short orbit. In \texttt{WB} and \texttt{DHC} coordinates the fractional difference in ZLK oscillation period to \texttt{IAS15} is $2.5\%$ and $28\%$, respectively. Both \texttt{TRACE} runs were ${\sim}20 \times$ faster than the \texttt{IAS15} run.

We conclude with some notes. First, the maximum ZLK eccentricity may grow due to octupole-order effects \citep[e.g.][]{naoz2013, li_eccentricity_2014}, making it much harder to map a maximum eccentricity to the initial binary inclination. Second, \texttt{WHFast} in Jacobi coordinates outperforms \texttt{IAS15} up to inclinations of $80^\circ$, or $e_\mathrm{max}\sim0.975$. However, for applications appropriate for \texttt{WHFast} it may be more natural to use a secular code.

\section{Conclusions}
\label{sec:conclusion}
We have implemented coordinates appropriate for integrations of planetary systems orbiting one component of a binary star system, originally derived by \cite{chambers2002wb}, into the hybrid integrator \texttt{TRACE}. Our implementation supports close encounters between pairs of planets, and between planets and either star in the binary system. We have conducted a parameter space study for when these coordinates offer significant benefits over DHC, and benchmarked the integrator's performance against other integrators in the \texttt{REBOUND} ecosystem.

Hybrid integrators have inherently limited and specific use cases. It is instructive to explicitly discuss when this integrator should be used over alternatives. We offer the following guidelines for choosing an integrator for planetary systems in \texttt{REBOUND}:
\begin{enumerate}
    \item If one is not limited by computation time, or exact trajectories are required, \texttt{IAS15} should be used.
    \item If computation time is a concern \textbf{and} there are no close encounters in a system, \texttt{WHFAST} should be used.
    \item If computation time is a concern, only qualitatively correct trajectories are required, \textbf{and} close encounters are expected, \texttt{TRACE} should be used. The choice of whether to use \texttt{TRACE} in the default \texttt{DHC} or \texttt{WB} coordinates then depends on the system's architecture. For systems about sun-like stars we offer the relation in Equation \ref{eq:plawfit} as a reasonable guideline for when \texttt{WB} coordinates are necessary.
\end{enumerate}
These use cases -- such as dynamical instabilities and planet formation in binary star systems -- are of great scientific value. For these use cases, we show that \texttt{TRACE} in \texttt{WB} coordinates offers sufficient accuracy without great computational costs. For large ensembles of chaotic systems, this may be the only viable integration method that is not prohibitively expensive.

\begin{acknowledgments}
T.L.\ is supported by a Flatiron Research Fellowship at the Flatiron Institute, a division of the Simons Foundation. This work has benefited from the use of the \texttt{rusty} cluster at the Flatiron Institute. We thank Hanno Rein, David Hernandez, Malena Rice, Vighnesh Nagpal, Max Goldberg, Yubo Su, Jared Goldberg, Quang Tran
and the members of the CCA Astronomical Data Group for helpful discussions.
\end{acknowledgments}

\software{\texttt{matplotlib} \citep{hunter_2007},
\texttt{REBOUND} \citep{Rein_2012}, \texttt{scipy} \citep{virtanen2020scipy}}

\bibliography{sample7}{}
\bibliographystyle{aasjournal}


\end{CJK*}
\end{document}